\documentclass[12pt]{article}

\usepackage{graphicx}
\begin{document}
\begin{center}
{\Large Spacetime Fluctuations and the Cosmological Constant Problem}
\vskip 0.3 true in
{\large J. W. Moffat}
\date{}
\vskip 0.3 true in
{\it Department of Physics, University of Toronto,
Toronto, Ontario M5S 1A7, Canada}

\end{center}

\begin{abstract}%
Spacetime geometry is treated as a fluctuating, stochastic quantum system allowing an
effective quantum gravity solution to the cosmological constant problem. A
Fokker-Planck equation for the probability density of spacetime metric fluctuations
is solved for an FRW universe and the most probable value for the metric scale
factor $R$ is for a very small cosmological constant.
\end{abstract}
\vskip 0.2 true
in e-mail: moffat@medb.physics.utoronto.ca

\section{Introduction}

The cosmological constant problem continues to defy all attempts to solve it in a
natural way that avoids any fine-tuning of
parameters~\cite{Weinberg,Witten,Straumann,Rugh}. Recent investigations of this
problem have focused on bulk-brane scenarios in which the standard model fields are
confined to a $(3+1)$ brane and gravity and moduli fields propagate in the
higher-dimensional bulk. The Randall-Sundrum models and various extensions of the
Rubakov-Shaposhnikov model have shown that a fine-tuning of brane
tension constants and the bulk cosmological constant cannot be
avoided~\cite{Rubakov}. Moreover the `self-tuning' models have also been confronted
with fine-tuning problems related to the existence of naked
singularities. These attempted solutions to the problem are based on
classical or semi-classical models and the issue of quantum gravity and other
quantum field theory vacuum corrections remains to be resolved. It is not yet
understood how string theory and D-brane theory can solve the cosmological constant
problem~\cite{Witten}. The supersymmetry solution to the cosmological constant
problem is prevented from being applicable, because supersymmetry must be broken for
energies below 1 TeV, and the fine-tuning of vacuum contributions to the cosmological
constant occurs already at the standard model energy scale, $E_{SM}\sim 10^2$ GeV,
and lower energies.

An early attempt by Hawking~\cite{Baum,Hawking} to solve the cosmological
constant problem was based on quantum cosmology and the existence of a four-index
tensor in four-dimensional spacetime. The wave function of the universe led to a
functional integral in Euclidean space with a probability distribution for the wave
function of the universe that favours a vanishingly small cosmological constant.
Coleman's wormhole
solution is also based on the probabilistic interpretation that the universe with
the smallest cosmological constant is the most probable~\cite{Coleman}. However, as
in the case of Hawking's proposal, it suffers from the difficulty of interpreting
Euclidean space gravity and other technical problems. The anthropic principle
solution~\cite{Weinberg2} is another attempt to resolve the problem, but it does not
explain how to solve the vanishingly small cosmological constant given the current
theory of particle physics and gravitation and its parameters.

Because we do not yet possess a satisfactory quantum gravity theory, the problem of
how to treat the linkage between general relativity (GR) and quantum
field theory remains somewhat in the dark. The standard semi-classical
approach, in which we write Einstein's field equations as
\begin{equation}
G_{\mu\nu}-\Lambda_0g_{\mu\nu}=8\pi G\langle T_{\mu\nu}\rangle,
\end{equation}
where $G_{\mu\nu}=R_{\mu\nu}-\frac{1}{2}g_{\mu\nu}R$, $\Lambda_0$ is the
`bare' Einstein cosmological constant and $\langle T_{\mu\nu}\rangle$ denotes the
expectation value of the stress-energy tensor treated as a quantum operator, is
basically inconsistent, for the left-hand side of the equation can only be
reconciled with the right-hand side by making the gravitational field change in a
discontinuous, acausal manner, a behavior unacceptable from the classical point of
view.

An `effective' quantum gravity interpretation of spacetime can be developed, in
which the metric tensor $g_{\mu\nu}$ is treated as a stochastic random variable,
which is defined by a set of possible values and a probability distribution $P(g)$
over this set with $P(g) \ge 0$~\cite{Moffat,Moffat2}. A $(3+1)$ foliation of
spacetime in GR is combined with the canonical formalism to derive a Langevin
equation for the conjugate canonical momentum, which is treated as a random
variable. In this formalism, both the gravitational field and the matter or vacuum
are treated in a probabilistic fashion, so that we have
\begin{equation}
\langle G_{\mu\nu}\rangle-\Lambda_0\langle g_{\mu\nu}\rangle=8\pi G\langle
T_{\mu\nu}\rangle. \end{equation} Thus, there is now no conflict between the two
sides of the equation.

By converting the Friedmann equation for a Friedmann-Robertson-Walker (FRW)
cosmology into a first-order stochastic differential equation, we can calculate the
approximate stationary probability density for the cosmological scale factor $R$,
and we find that the most probable value for $R$ in our universe occurs for a
very small cosmological constant.

\section{Stochastic Differential Equation for the Expansion of the Universe}

Wheeler and his co-workers~\cite{Wheeler} suggested that quantum fluctuations in the
metric tensor of spacetime, $g_{\mu\nu}$, should occur at the order of the Planck
length: $L_P=({\hbar}G/c^3)^{1/2}=1.6\times 10^{-33}{\rm cm}$. These fluctuations are
of order
\begin{equation}
\Delta g\sim\frac{L_P}{L},
\end{equation}
where $L$ is the spatial dimension of the region in a local Lorentz frame of reference.
These fluctuations are expected to be significant at the Planck scale $\sim
10^{-33}$ cm.

Let us consider a Friedmann-Robertson-Walker (FRW) spacetime with the metric
\begin{equation}
ds^2=dt^2-R^2(t)\biggl[\frac{dr^2}{1-kr^2}+r^2d\theta^2+r^2\sin^2\theta
d\phi^2\biggr],
\end{equation}
where $k=0,1,-1$ corresponding to a flat, closed and open universe, respectively. The
Einstein field equations are (c=1):
\begin{equation}
\label{Friedmann}
{\dot R}^2+k=\frac{8\pi G}{3}\rho R^2+\frac{\Lambda_0}{3}R^2,
\end{equation}
\begin{equation}
2R{\ddot R}+{\dot R}^2+k=-8\pi G pR^2+\Lambda_0 R^2,
\end{equation}
where ${\dot R}=\partial R/\partial t$.

From Eq.(\ref{Friedmann}), we can obtain the first order equation
\begin{equation}
\label{firstorderFriedmann}
{\dot R}=\biggl(\frac{8\pi G}{3}\rho R^2+\frac{\Lambda_0}{3}R^2-k\biggr)^{1/2},
\end{equation}
where we have chosen the positive square root for an expanding universe.
Including geometrical quantum ramdomness in the gravitational equations, leads to the
Langevin stochastic equation for the cosmic scale factor
$R$~\cite{Oksendal,Horsthemke}:
\begin{equation}
\label{stochReq}
dR_t=f(R_t)dt+\sigma\xi_t h(R_t)dt,
\end{equation}
where $R_t$ denotes the random scale factor, $\sigma$ denotes the intensity of the
quantum gravity fluctuations and $\xi_t$ is Gaussian fluctuations with
$\langle\xi_t\rangle=0$. We can take into account the fluctuations of the spacetime
geometry, in an effective way, by replacing the Planck length $L_P=\sqrt{G}$ by
a stationary random variable~\cite{Moffat}:
\begin{equation}
\label{RandomG}
L_{tP}\equiv \sqrt{G_t}=\sqrt{G}+\sigma\xi_t,
\end{equation}
where $\sqrt{G}$ denotes the mean value.

The extreme irregularity of Gaussian white noise means that the time derivative of
$R_t$ is not well defined. However, the standard theory of stochastic processes can
handle this difficulty by defining the equivalent integral equation
\begin{equation}
\label{integraleq}
R_t=R_0+\int^t_{t_0}f(R_y)dy+\sigma\int^t_{t_0}h(R_y)\xi_ydy.
\end{equation}
Two definitions of the stochastic integral $\int h(R_y)\xi_ydy$ have been given by
Ito~\cite{Ito} and Stratonovich~\cite{Stratonovich}.
Both the Ito and the Stratonovich definitions of the integral lead to a consistent
calculus.
The transition probability density $p(R,t)$ of the scale factor diffusion
process $R_t$ satisfies the Fokker-Planck equation (FPE):
\begin{equation}
\partial_tp(R,t)=-\partial_Rf(R,t)p(R,t)+\frac{\sigma^2}{2}\partial^2_R[h^2(R,t)p(R,t)].
\end{equation}

An important and useful solution of the FPEs can be obtained for stationary random
metric processes. We expect that a quantum gravitational system subjected to metric
fluctuations for a sufficiently long time will settle down to a stationary behavior.
This means that as time approaches infinity, the system will reach a state for which
the probability density $p_S(R)$ has a shape that does not change with time i.e. the
probability to find the system in the neighborhood of a particular state becomes time
independent. However, the sample paths $R_t(\omega)$ will in general not approach a
steady-state value $R_s(\omega)$, so that the state of the system still fluctuates.
However, these fluctuations are such that $R_t$ and $R_{t+\tau}$ have the same
probability density, $p_S(R)$.

The stationary solution $p_S(R)$ of the FPE satisfies
\begin{equation}
\partial_tp(R,t\vert R_i,t_i)+\partial_RJ(R,t\vert R_i,t_i)=0,
\end{equation}
where $R_i$ and $t_i$ denote the initial values of $R$ and $t$, respectively, and
$J$ is the probability flow current
\begin{equation}
J(R,t\vert R_i,t_i)=f(R)p(R,t\vert R_i,t_i)-\frac{\sigma^2}{2}
\partial_R[h^2(R)p(R,t\vert R_i,t_i)].
\end{equation}
The stationary FPE is then given by
\begin{equation}
\partial_RJ_S(R)=0,
\end{equation}
which implies that $J_S(R)={\rm constant}$ for $R\in[R_1,R_2]$. In the stationary
case, there are no sources or sinks in the universe for the probability. We have the
equation for the probability density $p_S(R)$:
\begin{equation}
-f(R)p_S(R)+\frac{\sigma^2}{2}\partial_Rh^2(R)p_S(R)=-J.
\end{equation}
We define the auxiliary function $u(R)=h^2(R)p_S(R)$ and obtain the solution to the
stationary FPE:
\begin{equation}
p_S(R)=\biggl(\frac{N}{h^2(R)}\biggr)\exp\biggl(\frac{2}{\sigma^2}
\int_{R_i}^R\frac{f(x)}{h^2(x)}dx\biggr)
$$ $$
-\biggl(\frac{2}{\sigma^2h^2(R)}\biggr)J\int_{R_i}^R\exp\biggl(\frac{2}{\sigma^2}
\int_y^R\frac{f(x)}{h^2(x)}dx\biggr)dy.
\end{equation}

We have $J=0$ when there is no flow of probability out of the universe.
By using the Ito prescription, we get
\begin{equation}
p_S(R)=\biggl(\frac{N}{h^2(R)}\biggr)\exp\biggl(\frac{2}{\sigma^2}\int_{R_i}^R\frac{f(x)}
{h^2(x)}dx\biggr).
\end{equation}
In this case, the normalization constant $N$ is
given by
\begin{equation}
N^{-1}=\int_{R_1}^{R_2}\frac{dR}{h^2(R)}\exp\biggl(\frac{2}{\sigma^2}
\int_{R_i}^R\frac{f(x)}{h^2(x)}dx\biggr) <\infty.
\end{equation}

\section{Metric Fluctuations and the Cosmological Constant}

Let us now apply our effective, stochastic quantum gravity formalism to the
cosmological constant problem.
We shall assume that matter is dominated by the vacuum density $\rho_{\rm vac}$ with
the equation of state: $\rho=-p=\rho_{\rm vac}={\rm constant}$ and that the universe
is spatially flat $k=0$. Then, we have from Eq. (\ref{firstorderFriedmann}):
\begin{equation}
\label{Requation}
dR=\biggl(\frac{8\pi}{3}\biggr)^{1/2}\sqrt{G}{\bar\rho}_{\rm vac}^{1/2}Rdt,
\end{equation}
where
\begin{equation}
\label{effectivevac}
{\bar\rho}_{\rm vac}=\rho_{\rm vac}+\frac{\Lambda_0}{8\pi G}=\frac{\Lambda}{8\pi G}.
\end{equation}
We obtain for $R$ the classical de Sitter solution
\begin{equation}
R=R_i\exp\biggl(\sqrt{\frac{\Lambda}{3}}t\biggr),
\end{equation}
where $R_i$ is the initial value of $R$ at $t=0$.

By replacing $R$ and $\sqrt{G}$ in Eq.(\ref{Requation}) by the random variables
$R_t$ and $\sqrt{G_t}$, we obtain the stochastic differential equation
(\ref{stochReq}) with
\begin{equation}
f(R_t)=\biggl(\frac{8\pi}{3}\biggr)^{1/2}\sqrt{G}{\bar\rho}^{1/2}_{\rm vac}R_t,
\end{equation}
\begin{equation}
h(R_t)=\biggl(\frac{8\pi}{3}\biggr)^{1/2}{\bar\rho}^{1/2}_{\rm vac}R_t.
\end{equation}
The approximate stationary probability density is given by
\begin{equation}
\label{stationaryprob}
p_S(R)=N\biggl(\frac{3GR_i^2}{\Lambda
R^2}\biggr)\exp\biggl[\biggl(\frac{2
G}{\sigma^2}\biggr)\sqrt{\frac{3}{\Lambda}}\ln(R/R_i)\biggr]
$$ $$
=N\biggl(\frac{3G}{\Lambda}
\biggr)\biggl(\frac{R}{R_i}\biggr)^{\beta-2},
\end{equation}
where $\beta=\biggl(\frac{2G}{\sigma^2}\biggr)\sqrt{\frac{3}{\Lambda}}$ and we have
used Eq.(\ref{effectivevac}) to define $\Lambda$.

We arrive at the result that the maximum
probability for the scale factor $R$, for a fixed value of the fluctuation
intensity $\sigma$ and $R/R_i > 1$, occurs for a very small cosmological constant
$\Lambda$. We illustrate this result in Fig 1., where we have plotted the stationary
probability density $p_S$ as a function of $R$ and $\Lambda$ for an arbitrary
normalization and for decreasing values of $\Lambda$. \vskip 0.2 in
\begin{center}\includegraphics[width=3.0in,height=3.0in]{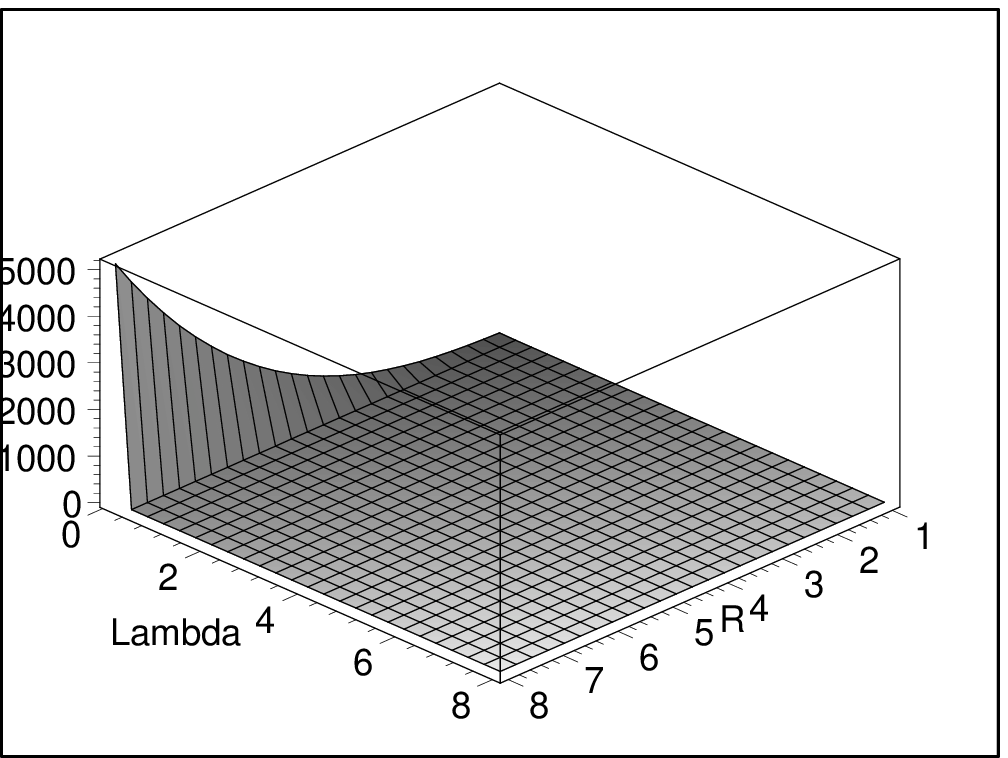}
\end{center} \vskip 0.1 in \begin{center}Fig. 1. Three-dimensional plot of
stationary probability density $p_S$ versus $R$ and $\Lambda$ for arbitrary
normalization and dimensionless units. \end{center} \vskip 0.1 true in

\section{Discussion of Results and Conclusions}

It is important to recognize that the cosmological constant $\Lambda$ is the true
effective cosmological constant, that would be measured in gravitational phenomena at
long ranges. It includes all effects of fields and any effects due to phase
transitions, such as those that occur at the electroweak energy scale, $10^2$ Gev,
and the QCD energy scale. We see that $p_S$ peaks sharply as $\Lambda$ decreases for
fixed values of the fluctuation intensity $\sigma$ and for $R/R_i > 1$. Thus, we
conclude that the most likely value of $R$ for all possible universes occurs for a
very small cosmological constant.

The intensity of the vacuum fluctuations of spacetime are expected to be
strong at the Planck energy $\sim 10^{19}$ Gev, where quantum gravitational effects
begin to dominate. However, this is the energy scale at which most models of
inflation begin to inflate~\cite{Brandenberger}. Although the quantum gravity
fluctuations lead to a small probable value of the cosmological constant in the
early expansion of the universe, this contradicts the usual expectation that it is
huge during the inflation period. This just underscores the cosmological constant
problem faced by scalar field, potential-driven models of inflation, namely, the need
for a very large cosmological constant initially in the universe, and the subsequent
observationally small bound on the constant in the present universe:
\begin{equation}
\vert\Lambda\vert \leq 10^{-56}\,{\rm cm}^{-2}\sim 10^{-84}\, {\rm GeV}^2.
\end{equation}
For the vacuum energy, the bound is
\begin{equation}
\rho_{\rm vac}\leq 10^{-29}\,{\rm g}/{\rm cm}^3\sim 10^{-47}\, {\rm GeV}^4\sim
10^{-9}\, {\rm ergs}/{\rm cm}^3.
\end{equation}
In our scenario, this would
presuppose that to uphold a large value of the cosmological constant (vacuum energy)
would require that spacetime is very smooth at the beginning of the universe with
$\sigma\sim 0$ in contradiction with our intuitive expectations that quantum gravity
effects should be very strong at the Planck energy scale.
If the vacuum energy in the universe is truly constant, as is demanded for ordinary
matter by the Bianchi identity: $\nabla_\nu T^{\mu\nu}\equiv 0$ and $\nabla_\nu
g^{\mu\nu}=0$ in Einstein's field equations, then when this constant is required to
be zero at the Planck energy scale, it should remain zero for the rest of the
universe's evolution. This scenario would avoid the embarrassing situation, in which
we have to somehow guarantee that $\Lambda$ relaxes to zero (or almost zero) whenever
there is a new phase transition as the universe evolves in time. If there exists
exotic `quintessence' matter~\cite{Perlmutter}, then a time dependent `cosmological
constant' could exist and the aforementioned difficulty would prevail. Moreover, we
are also faced in this case with the difficulty of explaining the coincidence of the
quintessence dark energy density with the present value of the matter energy density
(including cold dark matter). At present, the acceleration of the expansion
of the universe detected by supernovae data does not exclude a fit to all the current
observational data with a constant $\Lambda$.

Our explanation of the smallness of the cosmological constant is based on
an effective quantum gravity, stochastic argument, in which a probability density is
peaked at a vanishingly small cosmological constant. The result is similar in spirit
to the Baum, Hawking and Coleman explanations~\cite{Baum, Hawking,Coleman} (for a
detailed critique of these papers, see ref.~\cite{Weinberg}).
These proposals are based on a solution of the wave function of the universe $\Psi$
as an Euclidean path integral
\begin{equation}
\Psi\propto \int[dg][d\Phi]\exp(-S[g,\Phi]),
\end{equation}
where $S$ is the Einstein-Hilbert gravity action including matter terms, and the
integral is over all Euclidean 4-metrics $g$ and matter fields $\Phi$ defined on a
4-manifold. Hawking uses a three index field $B_{\mu\nu\sigma}$ and Coleman uses
wormholes as topological structures to obtain an
effective cosmological constant. There have been several important criticisms of
the underlying formalism, including the problem of having to use a complex
transformation to an Euclidean spacetime geometry, in which the Euclidean action
$S$ is unbounded from below. In our effective quantum gravity formalism, we
have not introduced any new fields beyond the standard Einstein-Hilbert action, but
we have based our probability analysis on a phenomenological stochastic formalism,
using a solution of the Fokker-Planck equation to obtain our probability
distribution. This formalism is well-founded in many applications to fluctuating
physical phenomena, and we suppose that it also applies to metric spacetime
fluctuations at the Planck energy scale.

The quantum gravity, stochastic fluctuations can be considered as the
fluctuations of micro-objects, e.g. strings or D-branes, which are expected to
dominate spacetime physics at the Planck energy. In this way, our stochastic,
effective quantum gravity formalism can be considered as a phenomenological picture
of how string and D-brane theory could resolve the cosmological constant problem.
However, the problem of supersymmetry breaking of string and D-brane theory would
have to be fully understood in such a scenario.

\vskip 0.3 true in

{\bf Acknowledgments}
\vskip 0.2 true in
I thank Michael Clayton for helpful discussions. This work was supported by the
Natural Sciences and Engineering Research Council of Canada.
\vskip 0.5 true in


\begin{thebibliography}{100}
\bibitem{Moffat2} In the preceding
reference [9], Eq. (165) should be replaced by
\begin{equation}
p_S(R)\sim C\biggl(\frac{3R^2}{8\pi\rho_{0,r}}\biggr)\exp\biggl[\biggl(\frac{2}
{\sigma^2}\biggr)\biggl(\frac{3G}{32\pi}\biggr)^{1/2}\biggl(\frac{R^2}{\rho_{0,r}^{1/2}}
\biggr)\biggr].
\end{equation} The conclusions regarding the smearing of the big bang
singularity remain the same.
\end{thebibliography}
\end{document}